\documentclass{article}
\usepackage[T1]{fontenc} 
\usepackage[utf8]{inputenc} 
\usepackage{amsfonts}
\usepackage{ismir,amsmath,cite,url}
\usepackage{graphicx}
\usepackage{color}
\usepackage{multirow}
\usepackage{mathtools}
\usepackage{lineno}
\usepackage{siunitx}
\usepackage{amssymb}
\usepackage{pifont}
\newcommand\blfootnote[1]{%
  \begingroup
  \renewcommand\thefootnote{}\footnote{#1}%
  \addtocounter{footnote}{-1}%
  \endgroup
}

\title{Towards Improving Harmonic Sensitivity and Prediction Stability for Singing Melody Extraction}

\multauthor
{Keren Shao* \hspace{1cm} Ke Chen* \hspace{1cm} Taylor Berg-Kirkpatrick \hspace{1cm} Shlomo Dubnov \vspace{0.1cm}} { 
 University of California San Diego \\
{\tt\small \{k5shao, knutchen, tberg, sdubnov\}@ucsd.edu}
}

\def\authorname{K. Shao, K. Chen, T. Berg-Kirkpatrick, S. Dubnov}

\usepackage[bookmarks=false,pdfauthor={\authorname},pdfsubject={\papersubject},hidelinks]{hyperref}

\begin{document}

\maketitle
\begin{abstract}
In deep learning research, many melody extraction models rely on redesigning neural network architectures to improve performance. In this paper, we propose an input feature modification and a training objective modification based on two assumptions. First, harmonics in the spectrograms of audio data decay rapidly along the frequency axis. To enhance the model's sensitivity on the trailing harmonics, we modify the Combined Frequency and Periodicity (CFP) representation using discrete $z$-transform. Second, the vocal and non-vocal segments with extremely short duration are uncommon. To ensure a more stable melody contour, we design a differentiable loss function that prevents the model from predicting such segments. We apply these modifications to several models, including MSNet, FTANet, and a newly introduced model, PianoNet, modified from a piano transcription network. Our experimental results demonstrate that the proposed modifications are empirically effective for singing melody extraction.\blfootnote{* The first two authors have equal contribution.}\blfootnote{Code: \href{https://github.com/SmoothKen/KKNet}{https://github.com/SmoothKen/KKNet}}
\end{abstract}

\section{Introduction}\label{sec:introduction}

Singing melody extraction is a challenging task that aims to detect and identify the fundamental frequency (F0) of singing voice in polyphonic music recordings. This task is more complicated than the monophonic pitch detection task due to the presence of various instrumental accompaniments and background noises, making it more difficult to accurately extract the singing melody. Singing melody extraction is not only crucial for music analysis by itself, but also has many downstream applications, such as cover song identification \cite{bytecover}, singing evaluation \cite{singingevaluation}, and music recommendation \cite{musrecom}.

Deep neural networks have been widely adopted in the singing melody extraction task to produce promising performance in terms of extraction accuracy. Early models \cite{mcdnn, patch-cnn, kim2018crepe} simply leveraged deep neural networks (DNN) and convolutional neural networks (CNN) \cite{cnn}. In more recent models, musical and structural priors were incorporated to improve performance. These include MSNet \cite{msnet} with a vocal detection component at the encoder-decoder bottleneck, joint detection model \cite{jdc} setting up an auxiliary network, and TONet \cite{tonet} with tone-octave predictions. Additionally, models can capture frequency relationships better with multi-dilation \cite{mdmr}, cross-attention networks \cite{ftanet}, graph-based neural networks \cite{graphicnet}, or harmonic constant-Q transform (HCQT) \cite{dsm}.

One of our observations relates to the input representations of the models, which play an important role in affecting the extraction performance. Timbre, which is closely related to harmonics, is one of the key components that helps models distinguish the vocal from other instruments. When the CFP representation \cite{cfp} is chosen as the input representation, its second feature, the generalized cepstrum, allows the model to learn the strength of harmonics of any given fundamental frequency in a localized manner. However, in music, the harmonics of a single sound usually decays rapidly along the frequency axis (detail in section \ref{sec:zcfp}), which can pose a challenge for the model to distinguish sounds that only differ significantly at the trailing harmonics. 

The transformation from the spectrum to the generalized cepstrum in CFP is a Fourier transform, and hence mostly captures the first few peaks with large magnitudes. As a result, this representation is not helpful in sensing the trailing harmonics. This motivates us to apply a different transformation function that produces a generalized cepstrum with better harmonics sensitivity.

Another observation relates to the vocal detection component. Extremely short vocal segments surrounded by non-vocal regions, and vice versa, rarely occur since vocalists typically sing a melody for at least half a second or rest for at least a few hundred milliseconds. Threshold-based removal \cite{bittner2017pitch}, mean or median filtering \cite{salamon2012melody, rosenzweig2019detecting} and Viterbi-based smoothing \cite{mauch2014pyin, bosch2016melody} are frequently used to address the problem. When they are implemented alongside a network-based algorithm, however, the network remains unaware of our smoothing intention and configuration. To investigate whether such awareness can increase the prediction performance, we derive a differentiable loss component that specifically penalizes spurious short-term predictions of these kinds during training, thus potentially guiding the model to produce consistently stable predictions.

In this paper, we propose two techniques that attempt to improve the two concerns mentioned above, namely the harmonic sensitivity and the prediction stability of singing melody extraction models. Our contributions are as follows:
\begin{itemize}
    \item We propose to use exponentially growing sinusoids along the frequency axis to transform the spectrum into the generalized cepstrum of the CFP representation. This approach is equivalent to taking a $z$-transform instead of Fourier transform, which increases the harmonic sensitivity of the input.
    \item We design a differentiable loss function as part of the training objective to teach the network to avoid predicting unrealistically short sequences of vocal and non-vocal at the voice detection bin. 
    \item We evaluate our techniques by applying them on several melody extraction models. Additionally, we adapt PianoNet \cite{pianonet}, originally developed for piano transcription, into the melody extraction task. Experimental results demonstrate state-of-the-art performance of our improved models.

\end{itemize}

\begin{figure}[t]
    \centering
    \includegraphics[width=\columnwidth]{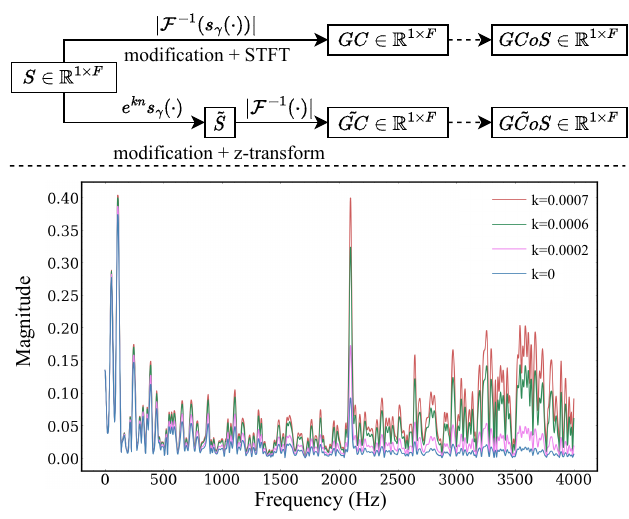}
    \vspace{-0.8cm}
    \caption{Top: the transformation pipeline of the original CFP representation, and our proposed $z$-CFP representation. Bottom: modified Spectrum $\tilde{S}$ with different growing rates $k$ applied. Note that the original CFP corresponds to the case of $k = 0$.}
    \label{fig:z-cfp}
    \vspace{-0.25cm}
\end{figure}



\section{Methodology}
In this section, we introduce three main parts of our methodology. First, we propose a modified CFP representation, $z$-CFP, to enhance the harmonic sensitivity of the network input. Second, we introduce extraction models used for evaluating our techniques, namely MSNet, FTANet, and PianoNet. Third, we propose a new loss function as part of training objective to improve the prediction stability of models. 

\subsection{$z$-CFP Representation for Harmonic Sensitivity} \label{sec:zcfp}
Our input representation of audio data is a modified version of the CFP representation. A CFP representation $X \in \mathbb{R}^{3 \times T \times F}$ contains three features, with $T$ the length of time frames and $F$ the number of frequency bins. \textbf{At each time slice}, it contains: (1) a power spectrum $S \in \mathbb{R}^{1 \times F}$; (2) a generalized cepstrum $GC \in \mathbb{R}^{1 \times F}$; and (3) a generalized cepstrum of spectrum $GCoS \in \mathbb{R}^{1 \times F}$, 

As illustrated in the upper part of Figure \ref{fig:z-cfp}, the standard CFP generation process begins by computing the frame-wise spectrum of an input audio waveform using short-time Fourier transform (STFT). We then obtain the magnitude of each spectrum, which serves as the first feature of CFP, denoted as $S$. To derive the second feature, we compute the generalized cepstrum using the following equation:

\begin{equation}
GC = |\mathcal{F}^{-1}(s_\gamma(S))| = |\mathcal{F}(s_\gamma(S))|
\end{equation}
where $\mathcal{F}$ and $\mathcal{F}^{-1}$ denotes the Fourier transform and its inverse, $s_\gamma: \mathbb{R} \to \mathbb{R}$ is an element-wise applied, logarithm-like modification function as described in \cite{cfp}, and the absolute value sign represents an element-wise complex norm operation. The second equality comes directly from the fact that norm of a complex number equals to that of its conjugate.

As mentioned in the introduction, $GC$ is not sensitive to the trailing harmonic dynamics, as it mostly captures the first few peaks with large magnitudes. Since the harmonics decay rapidly along with the frequency axis, we shall revert the decay to better preserve such dynamics. In other words, instead of applying complex sinusoids $\sum_n s_\gamma(S[n])e^{-iwn}$ as in Fourier transform ($n$ is the entry of frequency bins in $S$), we apply growing complex sinusoids $\sum_n s_\gamma(S[n])e^{(k-iw)n}$, where $k \in \mathbb{R}$ and $k > 0$. This is equivalent to taking a discrete $z$-transform $\sum_n s_\gamma(S[n])z^{-n}$, where $z = e^{iw-k}$.

In the actual implementation, $k$ is manually assigned and fixed across different $w$. Therefore, as illustrated in Figure \ref{fig:z-cfp}, we can separate the computation of $k$ part and $w$ part as follows:

\vspace{-0.5cm}
\begin{align}
\tilde{S}[n] &= e^{kn}s_\gamma(S[n]) \text{ for } \forall n \\
\tilde{GC} &= |\mathcal{F}^{-1}(\tilde{S})| = |\mathcal{F}(\tilde{S})|
\end{align}
\vspace{-0.5cm}

In the lower part of Figure \ref{fig:z-cfp}, we present $\tilde{S}$ of an audio waveform with different values of $k$. We can observe that the harmonics of $\tilde{S}$ at the tail gets amplified so that the subsequent Fourier transform can better capture their dynamics. While we observe some amplifications of harmonics at frequencies other than the fundamental frequencies, their magnitudes are always smaller than those of nearby fundamental frequencies. Therefore, they pose no sufficient distraction for the extraction model, as long as the chosen $k$ is not too large. In our experiments, we set $k=0.0006$.

We then generate the generalized cepstrum of spectrum $\tilde{GCoS}$ from cepstrum $\tilde{GC}$ the same way as in the original CFP. Finally, \textbf{each time slice} of our modified CFP representation $\tilde{X} \in \mathbb{R}^{3 \times T \times F}$ consists of $\{S, \tilde{GC}, \tilde{GCoS}\}$ with log-scaled frequency axis. For the rest of the paper, we denote it $z$-CFP.

\begin{figure}
    \centering
    \includegraphics[width=\columnwidth]{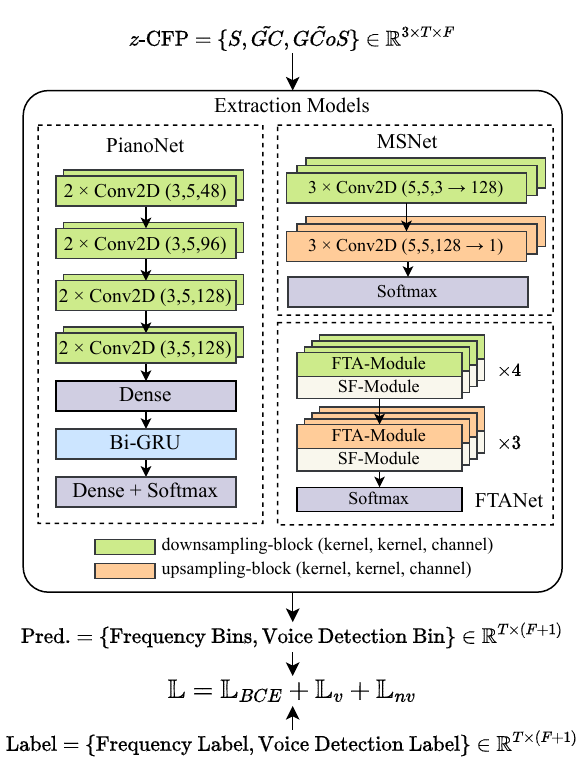}
    \vspace{-0.8cm}
    \caption{The model architecture. Note that we choose only one of the three extraction models at a time.}
    \label{fig:model-arch}
    \vspace{-0.5cm}
\end{figure}

\subsection{Model Architecture}
Our extraction models are referred from three state-of-the-art (SoTA) models, MSNet \cite{msnet}, FTANet \cite{ftanet}, and PianoNet \cite{pianonet}. Different from MSNet and FTANet, PianoNet is the SoTA model of piano transcription. Given its superior performance on piano transcription, we incorporate a sub-network of PianoNet into singing melody extraction, as we hypothesize that it may also yield good results for melody extraction.

MSNet contains a 3-layer encoder, a 3-layer decoder, and a bottleneck module. 
The channel size is shifted as $3 \rightarrow 32 \rightarrow 64 \rightarrow 128 \rightarrow 64 \rightarrow 32 \rightarrow 1$. 
The bottleneck module maps the encoder output to a 1-channel featuremap for voice detection. All 2D-convolutional layers come with $(5 \times 5)$ kernel size.
FTANet contains a 4-layer encoder, a 3-layer decoder, and a 4-layer bottleneck module. Both encoder and decoder contain FTA-modules and SF-modules to process the audio latent features. The channel size is shifted from $3$ to $128$, then back to $1$. More specifications of MSNet and FTANet can be found in their papers \cite{msnet, ftanet}. 


The PianoNet we use for this task is modified from a sub-network of \cite{pianonet}. It starts with four convolutional blocks, each block containing two 2D-convolutional layers with kernel sizes (3, 5) and (3, 3) respectively, a batch normalization layer and a ReLU activation. Then it is followed by bidirectional-GRU and softmax layers, with dense layers as transitions. The layer bias is turned off for all layers before the Bi-GRU.

Figure \ref{fig:model-arch} illustrates a more detailed structure of the three extraction models. Following the pipeline, we first process the audio waveform into $z$-CFP representations. Then we feed them into the extraction model, which produces output feature maps $\tilde{Y} \in \mathbb{R}^{T \times (F+1)}$. The additional one feature along the frequency axis denotes the voice detection bin output. It is then compared against the ground truth label $Y \in \mathbb{R}^{T \times (F+1)}$, through the loss function introduced in the following section.

\begin{figure}[t]
 \centering
 \includegraphics[width=\columnwidth]{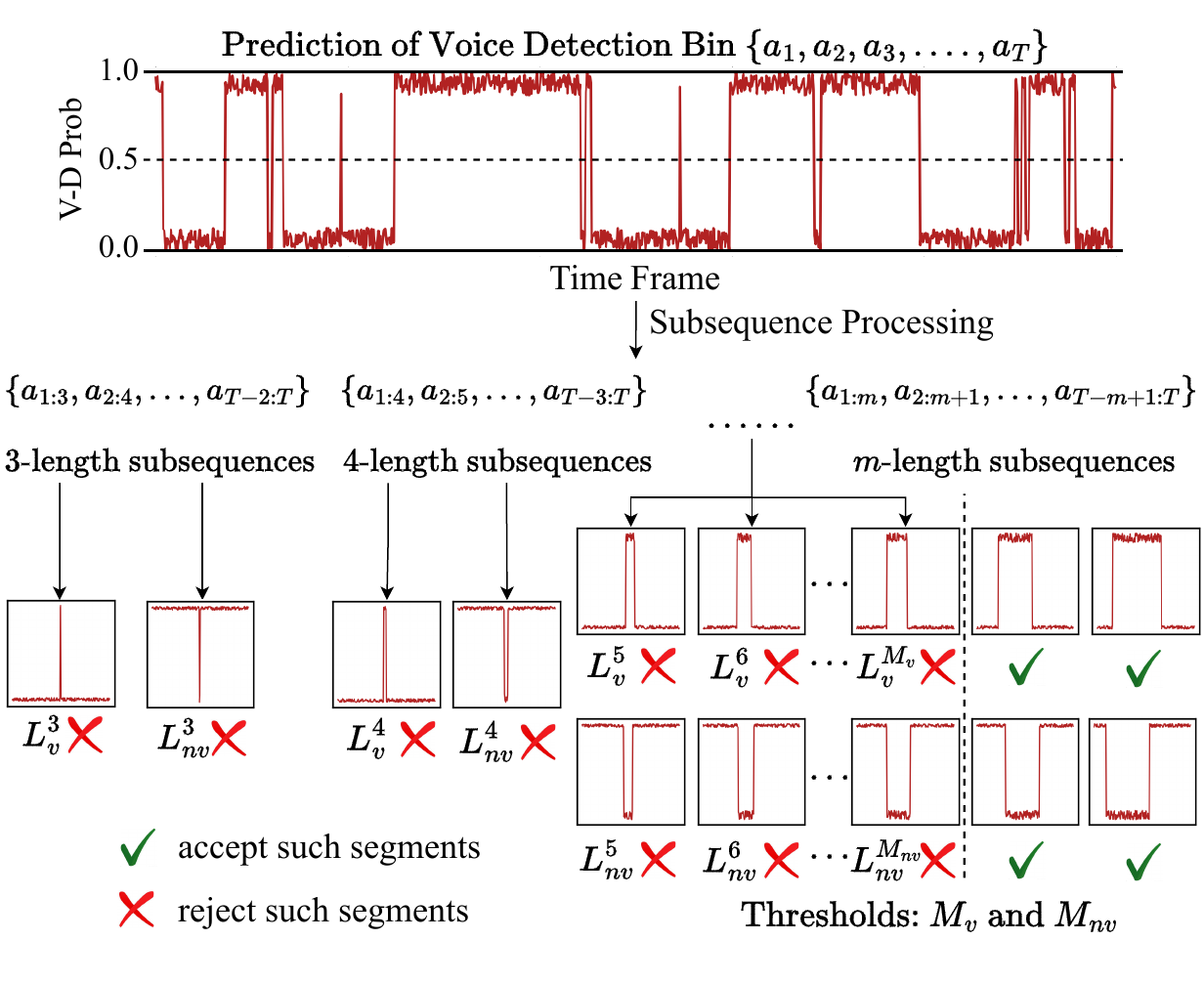}
 \vspace{-1cm}
 \caption{The illustration of how we perform the loss functions $\mathbb{L}_{v}$ and $\mathbb{L}_{nv}$ on the subsequences of the voice detection prediction. Each loss components $L$ are used to give large penalties (i.e., \ding{55}) to certain types of subsequences. }
 \label{fig:loss}
 \vspace{-0.5cm}
\end{figure}

\subsection{Loss Function for Prediction Stability}
We add two differentiable training objectives, $\mathbb{L}_{v}$ and $\mathbb{L}_{nv}$, to the conventional binary cross entropy loss $\mathbb{L}_{BCE}$ to teach the extraction model to avoid unrealistically short vocal and non-vocal sequences at \textbf{the vocal detection bin}. Since the design for these two cases are symmetric, we first introduce the loss object $\mathbb{L}_{v}$, for the vocal case.

As shown on the top of Figure \ref{fig:loss}, the predictions at the vocal detection bin is a time series $\{a_1,a_2,...,a_T\}$. First, since our training objectives are dealing with certain types of short burst segments of vocal and non-vocal, we extract all possible subsequences, with stride 1. For example, for 3-length subsequences we have $\{a_{1:3}, a_{2:4}, ..., a_{T-2:T}\}$, and similarly $\{a_{1:4}, a_{2:5}, ..., a_{T-3:T}\}$ for subsequences of length 4, etc.

Second, to simplify the problem a bit at the beginning, we assume that the voice detection output is binary valued $a \in \{0,1\}$. Formally, we do not want ``sharp-burst" sequences inside the following set:

\vspace{-0.25cm}
\begin{small}
\begin{align}
B_v = \bigcup_{m = 3}^{M_v} \{a_1...a_m | a_1 = a_m = 0, a_i = 1 \text{ for } \forall i \neq 1, m\}
\end{align}
\end{small}
where $M_v$ is a hyperparameter threshold, above which the duration of vocal segments becomes reasonable. Figure \ref{fig:loss} illustrates examples of ``sharp-burst" sequences in $B_v$ (and $B_{nv}$) as red segments inside black-border boxes.  

Suppose $m = 3$, all possible binary sequences are $\{000, 001, 010, 011, 100, 101, 110, 111\}$ and $010 \in B_v $. To make the model avoid predicting the short burst vocal segment, i.e., $010$, we construct a polynomial objective that can fulfill the goal by satisfying the following:

\begin{table*}[t]
\small
\renewcommand\arraystretch{1.3} 

\resizebox{\textwidth}{!}{
\setlength{\tabcolsep}{0.85mm}{
\centering
\begin{tabular}{l|ccccc|ccccc|ccccc}
\hline \hline
\multicolumn{1}{c|}{Dataset} & \multicolumn{5}{c|}{ADC 2004} & \multicolumn{5}{c|}{MIREX 05} & \multicolumn{5}{c}{MEDLEY DB} \\
\cline{3-5} \cline{8-10} \cline{13-15} 
\multicolumn{1}{c|}{Metrics} &  VR  & VFA$\downarrow$ & RPA & RCA &  OA & VR & VFA$\downarrow$ & RPA & RCA  & OA & VR & VFA$\downarrow$ & RPA & RCA  & OA  \\
\cline{1-2} \cline{3-16} 
PianoNet & 87.21 & 14.62 & 84.28 & 84.30 & 84.48 & 91.98 & 6.14 & 86.54 & 86.55 & 89.19 & 69.38 & 13.74 & 61.81 & 62.80 & 73.70 \\
PianoNet + $z$-CFP & 88.25 & \textbf{7.58} & 84.87 & 84.93 & 86.27 & \textbf{93.44} & 6.21 & 86.78 & 86.79 & 89.33 & 68.76 & \textbf{11.91} & 62.22 & 63.10 & \textbf{74.80} \\

PianoNet + 3 point median & 87.33 & 14.58 & 84.35 & 84.38 & 84.55 & 92.08 & 6.15 & 86.60 & 86.62 & 89.23 & 69.49 & 13.77 & 61.86 & 62.86 & 73.71 \\

PianoNet + 7 point median &
87.58 & 14.53 & 84.46 & 84.48 & 84.65 & 92.47 & 6.14 & 86.78 & 86.8 & 89.35 & 69.71 & 13.83 & 61.92 & 62.91 & 73.71 \\

PianoNet + 15 point median & 89.13 & 14.21 & 84.89 & 84.91 & 85.06 & 93.27 & 6.58 & 86.82 & 86.84 & 89.21 & 70.31 & 14.43 & 61.91 & 62.90 & 73.42 \\

PianoNet + $\{\mathbb{L}_v, \mathbb{L}_{nv}\}$ & \textbf{90.92} & 13.58 & \textbf{86.06} & \textbf{86.12} & 86.13 & 91.87 & \textbf{5.79} & 87.50 & 87.50 & \textbf{89.94} & \textbf{71.16} & 15.77 & \textbf{63.66} & \textbf{64.81} & 73.66 \\
PianoNet + $z$-CFP + $\{\mathbb{L}_v, \mathbb{L}_{nv}\}$ & 90.50 & 7.99 & 85.76 & 85.82 & \textbf{86.92} & 92.84 & 6.39 & \textbf{87.57} & \textbf{87.59} & 89.76 & 68.88 & 12.29 & 62.05 & 62.91 & 74.53 \\
\hline
MSNet & 89.78 & 23.12 & 80.83 & 81.60 & 80.10 & 84.85 & \textbf{11.44} & 77.76 & 78.09 & 81.68 & 53.49 & \textbf{9.41} & 46.90 & 48.24 & 68.15 \\
MSNet + $z$-CFP + $\{\mathbb{L}_v, \mathbb{L}_{nv}\}$ & \textbf{90.61} & \textbf{14.62} & \textbf{81.96} & \textbf{82.57} & \textbf{82.59} & \textbf{88.38} & 14.85 & \textbf{80.83} & \textbf{81.01} & \textbf{82.39} & \textbf{62.95} & 14.60 & \textbf{53.60} & \textbf{55.31} & \textbf{69.07} \\
\hline
FTANet & 81.26 & \textbf{2.70} & 77.17 & 77.36 & 80.89 & 87.34 & \textbf{5.11} & 81.56 & 81.61 & 86.40 & 62.44 & 10.41 & 55.94 & 56.58 & 72.30 \\
FTANet + $z$-CFP + $\{\mathbb{L}_v, \mathbb{L}_{nv}\}$ & \textbf{90.29} & 10.83 & \textbf{85.06} & \textbf{85.19} & \textbf{85.82} & \textbf{90.50} & 6.63 & \textbf{83.94} & \textbf{83.99} & \textbf{87.36} & \textbf{63.71} & \textbf{9.35} & \textbf{56.32} & \textbf{57.29} & \textbf{73.02} \\

\hline \hline

\end{tabular}}}
\caption{Ablation studies on ADC2004, MIREX05 and MedleyDB testsets. Baselines use CFP as the input representation and $\mathbb{L}_{BCE}$ as the loss function. $\{\mathbb{L}_v, \mathbb{L}_{nv}\}$ denotes the use of our proposed loss function in section 2.3. Among median filter sizes in the range $[3, 100] \subset \mathbb{Z}$, 3 point works best for MedleyDB, 7 point works best for MIREX 05, and 15 point works best for ADC 2004. But they neither significantly outperform our proposed loss component in any single dataset, nor uniformly outperform in all three datasets.}

\label{tab:aba_test}
\end{table*}

\vspace{-0.4cm}
\begin{small}
\begin{align}
L_v^3(a_1 a_2 a_3) = 
\left\{
\begin{matrix}
 1 &  \text{where} ~ a_1 a_2 a_3 = 010 \\ 
 0 & \text{otherwise}
\end{matrix}
\right.
\end{align}
\end{small}
A decent choice will then be 
\begin{small}
\begin{align}
    &L_v^3(a_1a_2a_3) = (1-a_1)a_2(1-a_3)
\end{align}
\end{small}
which can be easily extended to sequences with longer length $m$.

\vspace{-0.5cm}
\begin{small}
\begin{align}
    &L_v^4(a_1a_2a_3a_4) = (1-a_1)a_2a_3(1-a_4) \nonumber \\
    &~~~~~~~\vdotswithin{\ldots} \nonumber\\
    &L_v^m(a_1...a_m) = (1-a_1)(1-a_m)\prod_{i=2}^{m-1} a_i
\end{align}
\end{small}


However, there is a small caveat in this extension when we move back from binary values to probability values $a \in [0,1]$. For example, our loss component will be having trouble capturing sequences like $\{0.1, 0.4, 0.6, 0.1\}$ and $\{0.1, 0.6, 0.4, 0.1\}$ as both $L_v^3$ and $L_v^4$ result in relatively small values. However, we observe that polynomials $(1-a_1)(1-a_2)a_3(1-a_4)$ and $(1-a_1)a_2(1-a_3)(1-a_4)$ respectively works better than our original $L_v^4$, but still insufficient to work standalone.

Since none of the polynomials above gives high values to sequences outside of $B_v$ in $4$-length, a simple solution would be to redefine $L_v^4$ to be the sum of all such polynomials:

\vspace{-0.5cm}
\begin{small}
\begin{align}
    L_v^4 &= (1-a_1)(1-a_4)(a_2a_3 + a_2(1-a_3) + (1-a_2)a_3) \nonumber \\
    &~~~~~~~\vdotswithin{\ldots} \nonumber\\
    L_v^m &=  (1-a_1)(1-a_m) \sum_{\substack{c_1...c_m \in \{0,1\}^m \\ \text{ at least one } c_i \neq 0}} \prod_{i = 2}^{m-1} a_i^{c_i}(1-a_i)^{1-c_i} \nonumber \\
    &= (1-a_1)(1-a_m)(1 - \prod_{i = 2}^{m-1} (1 - a_i)) 
\end{align}
\end{small}

This redefined loss $L_v$ allows better recognition of the bad sequences mentioned above while not falsely flagging sequences outside of $B_v$. Furthermore, when dealing with longer sequences, for example $\{0.1, 0.9, ..., 0.9, 0.1\}$ with increasingly many $0.9$s in the middle, the original $L_v$'s output quickly diminishes while the redefined $L_v$ does not.

This redefined objective does come with a small side effect, as it over-counts the shorter bad sequences. For example, $(0.1, 0.9, 0.1, 0.1)$ now gets a high loss value not only in $L_v^3$, but also in $L_v^4$. However, we believe this side effect does not have significant impact as it does not matter whether neural network decides to stop producing shorter bad sequences or longer bad sequences first.

A further improvement is to pass the value of $L_v^m$ into the S-curve function:
\begin{align}
    L_v^m \leftarrow \frac{(L_v^m)^r}{(L_v^m)^r + (1-L_v^m)^r} 
\end{align}
where $r \in \mathbb{R}$ and $r > 1$. It will amplify those sequences that receive loss values closer to 1 and suppress those sequences with loss values closer to 0. 

Finally, for each $m \in [3, M_v]$, we compute $L_v^m$ across all $m$-length subsequences in the model's output. The aggregated loss function $\mathbb{L}_v$ is then computed by concatenating all these $L_v^m$ arrays and taking the average.

Now analogously, assuming non-vocal sequences beyond length $M_{nv}$ become reasonable, we can perform the same analysis on the following set of sequences:

\vspace{-0.5cm}
\begin{small}
\begin{align}
B_{nv} = \bigcup_{m = 3}^{M_{nv}} \{a_1...a_m | a_1 = a_m = 1, a_i = 0 \text{ for } \forall i \neq 1, m\}
\end{align}
\end{small}
and consequently obtain $\mathbb{L}_{nv}$. Practically, $L_{nv}^m$ of any sequence $a_1...a_m$ can be computed as $L_v^m$ of the flipped sequence $b_1...b_m$, where $b_i = 1 - a_i$ for all $i \in \{1..m\}$. Our final loss function will then be:
\begin{align}
    \mathbb{L} = \mathbb{L}_{BCE} + \mathbb{L}_{v} + \mathbb{L}_{nv}
\end{align}



\section{Experiments}

\subsection{Datasets and Experiment Setup}

For the training data, we complied with the setting of \cite{ftanet, tonet} and chose all \num{1000} Chinese pop songs from MIR-1K\footnote{http://mirlab.org/dataset/public/MIR-1K.zip} and 35 vocal tracks from MedleyDB \cite{medleydb}. For the testing data, we chose 12 tracks in ADC2004 and 9 tracks in MIREX05\footnote{https://labrosa.ee.columbia.edu/projects/melody/}. We also selected 12 tracks from MedleyDB that are disjoint from those already used for training.

For the signal processing part, we used \SI{8000}{\hertz} sampling rate to process audio tracks. We use a window size of \num{768}, a hop size of \num{80} to compute the STFT of audio tracks. Note that the time resolution of our labels is 0.01 seconds, and this hop size was chosen to match that. Then, when creating $z$-CFP representations, we set the time dimension of the representation to be $T = 128$, or \num{1.28} seconds, and the number of frequency bins $F=360$, or \num{60} bins per octave across \num{6} octaves. The start and stop frequencies are \SI{32.5}{\hertz} and \SI{2050}{\hertz}. Hence, the input shape becomes $X \in \mathbb{R}^{3 \times 128 \times 360}$ and the output/label shape becomes $Y \in \mathbb{R}^{128 \times 361}$. 

Within the extra loss component, we set the duration threshold of vocal segments $M_v=30$ (0.3 seconds), the duration threshold of non-vocal segments $M_{nv}=7$ (0.07 seconds), and the S-curve exponent parameter $r=5$.  

For the training hyperparameters, we use a batch size of \num{10}, the Adam optimizer \cite{adam} with a fixed learning rate of \num{1e-4}. The maximum training epoch is \num{500}. During the evaluation, we use the standard metrics of the singing melody extraction task, namely, voice recall (VR), voicing false alarm (VFA), raw pitch accuracy (RPA), raw chroma accuracy (RCA), and overall accuracy (OA) from the \texttt{mir\_eval} library \cite{mir_eval}. Following the convention of this task, overall accuracy (OA) is regarded as the most important metric. All models are trained and tested in NVIDIA RTX 2080Ti GPUs and implemented in PyTorch\footnote{https://pytorch.org/}.







\subsection{Ablation Study}
We choose three extraction models, namely MSNet \cite{msnet}, FTANet \cite{ftanet}, and PianoNet \cite{pianonet}, to evaluate our $z$-transform and loss functions. We conducted ablation studies and presented the results in Table \ref{tab:aba_test}. We re-trained these models from scratch, and the results are largely consistent with the original reports of \cite{msnet, ftanet, tonet}. The option $z$-transform denotes the use of $z$-CFP representations. Note that  $\{\mathbb{L}_v, \mathbb{L}_{nv}\}$ in the table denote the use of loss functions to address short burst segments of vocal and non-vocal. Due to the page limitation, we present a detailed ablation study on PianoNet while ablating MSNet and FTANet in an all-or-nothing fashion.


From Table \ref{tab:aba_test} we can clearly observe decent performance of both $z$-CFP and $\{\mathbb{L}_v, \mathbb{L}_{nv}\}$ when added to the PianoNet, MSNet, and FTANet. Among these results, the addition of loss functions $\{\mathbb{L}_v, \mathbb{L}_{nv}\}$ increases the overall accuracy while improving the VR, RPA, and RCA. The median filter postprocessing \cite{rosenzweig2019detecting} is used as a comparison. Since our loss component focuses on the vocal detection, we took the pitches predicted by median filters only when the original predictions are non-vocal. Further, to ensure fairness, we optimized the filter size against each single dataset within the range $[3, 100] \subset \mathbb{Z}$ and listed the evaluation results of those optimal ones. As we can see in Table 1, none of these median filters outperforms our loss component in a consistent manner, nor do they obtain considerable margins in any single dataset.

The $z$-CFP also increases several metrics, especially either VR or VFA, on each dataset. This indicates that by preserving more dynamics in the high frequency bins, the model can distinguish different sounds better and consequently improve the extraction performance. Also, note that unlike TONet \cite{tonet} and JDC \cite{jdc}, which achieved this through model design or music inductive bias, this technique relies solely on the inherent characteristics of the data.




When we incorporate both techniques into the extraction models, we observe a promising increase in each metric compared to the original models. However, we notice that some models with both techniques carried do not yield better performance than the models carrying only one of the techniques. These models appear to be an averaging weighting or an ensemble of models improved with either technique, implying better generalization.

\begin{table}[t]
\small
\renewcommand\arraystretch{1.2} 
\resizebox{\columnwidth}{!}{
\centering
\begin{tabular}{c|ccccc}
\hline \hline
Dataset & \multicolumn{5}{c}{ADC 2004} \\
\cline{2-6} 
Metrics &  VR  & VFA$\downarrow$ & RPA & RCA & OA \\
\hline
MCDNN \cite{mcdnn} & 65.0 & 10.5 & 61.6 & 63.1 & 66.4  \\
DSM \cite{dsm} & 89.2 & 51.3 & 75.4 & 77.6 & 69.8 \\
MSNet \cite{msnet} & 89.8 & 23.1 & 80.8 & 81.6 & 80.1 \\
FTANet \cite{ftanet} & 81.3 & \textbf{2.7} & 77.2 & 77.4 & 80.9 \\
TONet \cite{tonet} & \textbf{91.8} & 17.1 & 82.6 & 82.9 & 82.6 \\
SpecTNT* \cite{spectnt} & 85.4 & 8.2 & 83.5 & 83.6 & 85.0 \\
H-GNN* \cite{graphicnet} & 89.2 & 21.3 & 84.8 & 86.1 & 83.9 \\
\textbf{Ours} & 90.5 & 8.0 & \textbf{85.7} & \textbf{85.8} & \textbf{86.9} \\
\hline \hline
Dataset & \multicolumn{5}{c}{MIREX 05} \\
\cline{2-6} 
Metrics &  VR  & VFA$\downarrow$ & RPA & RCA & OA \\
\hline
MCDNN \cite{mcdnn} & 66.5 & \textbf{4.6} & 64.1 & 64.4 & 75.4\\
DSM \cite{dsm}  & 91.4 & 45.3 & 75.7 & 77.0 & 68.4 \\
MSNet \cite{msnet} & 84.8 & 11.4 & 77.8 & 78.1 & 81.7 \\
FTANet \cite{ftanet} & 87.3 & 5.1 & 81.6 & 81.6 & 86.4  \\
TONet \cite{tonet} & 91.6 & 8.5 & 83.8 & 84.0 & 86.6 \\
SpecTNT* \cite{spectnt} & 82.2 & 8.7 & 77.4 & 77.5 & 82.5 \\
H-GNN* \cite{graphicnet} & \textbf{93.2} & 21.7 & 85.2 & 86.4 & 81.3 \\
\textbf{Ours}  & 92.8 & 6.4 & \textbf{87.6} & \textbf{87.6} & \textbf{89.8} \\
\hline \hline
Dataset & \multicolumn{5}{c}{MEDLEY DB} \\
\cline{2-6} 
Metrics &  VR  & VFA$\downarrow$ & RPA & RCA & OA \\
\hline
MCDNN \cite{mcdnn} & 37.4 & \textbf{5.3} & 34.2 & 35.3 & 62.3 \\
DSM \cite{dsm} & \textbf{86.6} & 44.3 & \textbf{70.2} & \textbf{72.4} & 64.8 \\
MSNet \cite{msnet} & 53.5 & 9.4 & 46.9 & 48.2 & 68.1 \\
FTANet \cite{ftanet} & 62.4 & 10.4 & 55.9 & 56.6 & 72.3 \\
TONet \cite{tonet} & 64.2 & 12.5 & 56.6 & 58.0 & 71.6 \\
SpecTNT* \cite{spectnt} & 62.7 & 18.8 & 54.7 & 56.4 & 63.9 \\
H-GNN* \cite{graphicnet} & 71.7 & 21.6 & 61.2 & 65.8 & 67.9 \\
\textbf{Ours} & 68.9 & 12.3 & 62.1 & 62.9 & \textbf{74.5} \\ 
\hline \hline
\end{tabular}}
\caption {The comprehensive performance comparison among our improved models and current baselines. }  
\label{tab:exp_result}
\vspace{-0.4cm}
\end{table}
%








\subsection{Comprehensive Performance Comparison} \label{sec:cpc}
Table \ref{tab:exp_result} presents the results as we compare our best model, i.e., PianoNet with $z$-transform and $\{\mathbb{L}_v, \mathbb{L}_{nv}\}$, with other SoTA models. Among these SoTAs, there are two models with ``*", indicating that these are only partial comparisons. For SpecTNT \cite{spectnt}, since there is no official open-source implementation, we report its results based on our own re-implementation. For H-GNN \cite{graphicnet}, we directly copied its reported performance from the original paper. 

From Table \ref{tab:exp_result}, our improved PianoNet with $z$-transform and $\{\mathbb{L}_v, \mathbb{L}_{nv}\}$ yield the best OA performance over all datasets, the best RPA and RCA on ADC 2004 and MIREX 05 datasets. We do note, despite the use of the extra loss component, that our model's VFA is not necessarily the smallest. This is because the extra loss component only targets a particular type of false positive, and is not meant to minimize the false positive rate in general. For example, sometimes the network's vocal to non-vocal transition happens later than the reference labels. In this case, since the vocal sequence itself lasts long enough, the extra loss component will not mark this type of false positives. Addressing this type of errors is potentially a future work.

Another thing we found is that the PianoNet, as one of SoTAs in the piano transcription task and ported by us to the melody extraction task in this paper, has already yields very high performance on MIREX 05 dataset. This indicates that there may exist more powerful network architectures for this task yet to be explored. Additionally, it is noteworthy that our proposed PianoNet architecture has a small number of parameters (5.5 million), which is comparable with MCDNN (5.6 million), FTANet (3.4 million) and far less than TONet (152 million). This demonstrates its potential in practical applications where computational resources are limited. Again, as demonstrated in Table \ref{tab:aba_test}, our techniques could help models other than PianoNet achieve higher performance than their original versions.

\begin{figure}[t]
 \centering
 \includegraphics[width=\columnwidth]{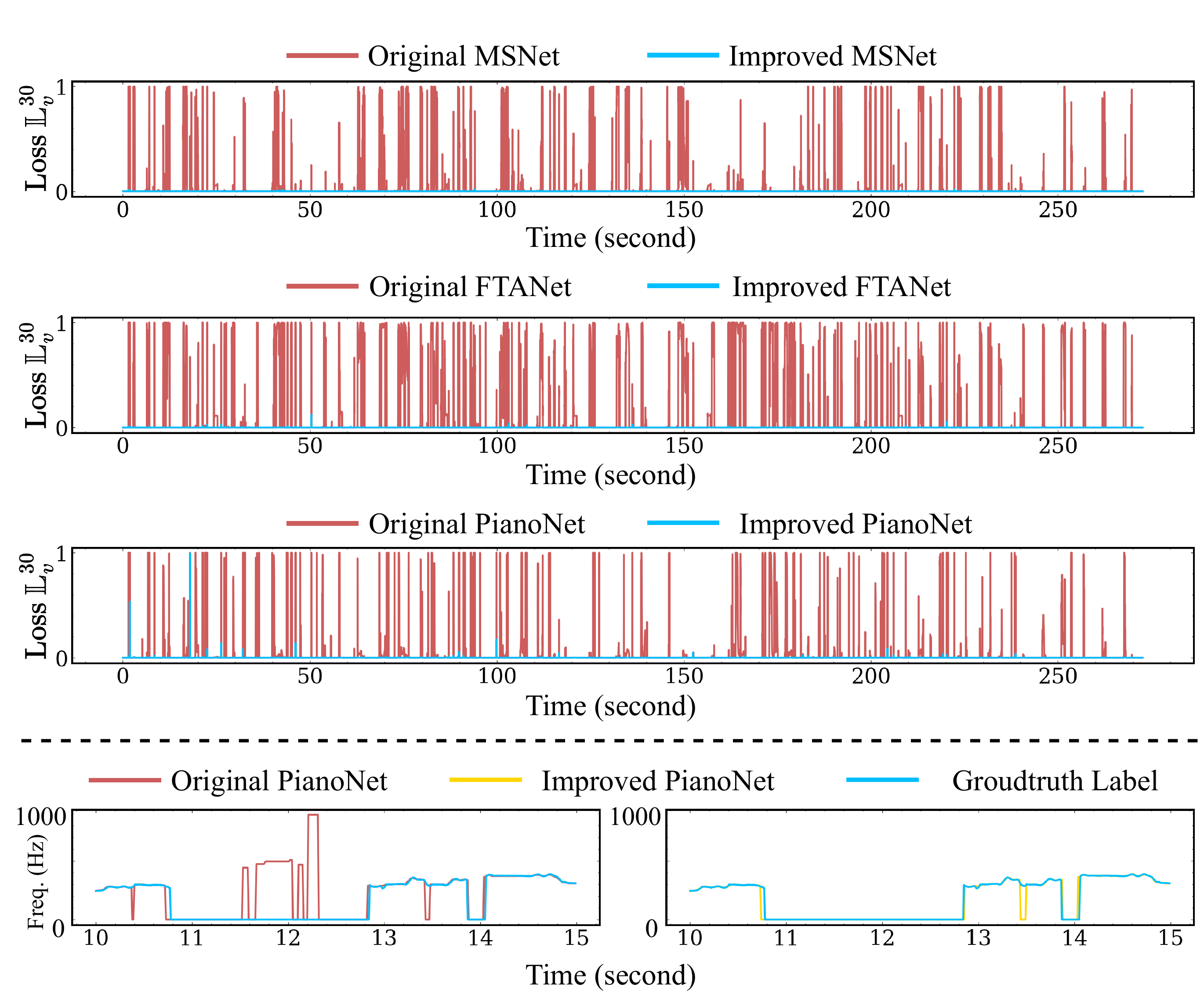}
 \caption{The effect of applying the loss $\mathbb{L}_{v}$ and $\mathbb{L}_{nv}$. The top three plots are values of $L_{v}^{30}$ across the entire MIREX05 dataset. The bottom two plots are one 5-sec MIREX05 predictions.}
 \label{fig:visualization}
 \vspace{-0.4cm}
\end{figure}

\subsection{Loss Value and Extraction Visualization} \label{sec:visualization}
To empirically verify if applying the polynomial loss functions $\mathbb{L}_v$ and $\mathbb{L}_{nv}$ could reduce the voice detection errors, i.e., short burst segments of vocal and non-vocal, we visualize two types of plots in Figure \ref{fig:visualization}. The top three plots demonstrate the loss values of $L_v^{30}$ between the original extraction models and the improved models with $\mathbb{L}_v$ and $\mathbb{L}_{nv}$, across the entire MIREX05 dataset (i.e., we concatenate all tracks in the dataset). We see that cases in which the improved models' prediction receive loss values close to 1 diminishes comparing to those of the original models. This phenomenon implies that after applying $\mathbb{L}_v$ and $\mathbb{L}_{nv}$, the chance of models to predict short burst segments significantly reduces.

The pair of plots in the last row compares the prediction performance of PianoNets, trained without and with the extra loss components, on a zoomed-in section of MIREX05. Note that the original PianoNet has a short burst non-vocal segment in between the 10th second and 11th second. Further, it has a considerable number of short burst vocal segments around the 12th second. Once trained with the extra loss components, these issues are resolved. 
Also note that both the original version and the improved version make a mistake in between the 13th and the 14th second. This is because the length of that non-vocal transition is greater than our threshold $M_{nv}$, which ends up not triggering $\mathbb{L}_{nv}$. All these observations further verify the effectiveness of our proposed loss components.
%
\vspace{-0.5cm}
\section{Conclusion}

We propose two techniques to respectively utilize the two assumptions for improving the singing melody extraction performance. The use of $z$-transform in generating cepstrum allows the network to better recognize the strength of harmonics of any fundamental frequencies. Empirically, while the trailing harmonics of those frequencies that do not actually appear in the audio also get elevated, the benefit of the technique is greater than its setback. Our extra loss components make the network less prone to predict vocal and non-vocal sequences are unreasonably short, while not affecting the network's overall accuracy due to its differentiability. We regard these two techniques as decent improvements on singing melody extraction models.

\section{Acknowledgments}
We would like to thank the Institute for Research and Coordination in Acoustics and Music (IRCAM) and Project REACH: Raising Co-creativity in Cyber-Human Musicianship for supporting this project. This project has received funding from the European Research Council (ERC REACH) under the European Union's Horizon 2020 research and innovation programme (Grant Agreement \#883313).

\bibliography{refs}

%
%
%
%
%

\end{document}